# COPYRIGHT AND PROMOTION: OXYMORON OR OPPORTUNITY?

**Teresa Numerico, Jonathan P. Bowen**
London South Bank University, Institute for Computing Research
Faculty of BCIM, Borough Road, London SE1 0AA, UK
{teresa.numerico, jonathan.bowen}@lsbu.ac.uk
http://www.lsbu.ac.uk/bcimicr & http://www.jpbowen.com

**Abstract**
**Copyright in the cultural sphere can act as a barrier to the dissemination of high-quality information. On the other hand it protects works of art that might not be made available otherwise. This dichotomy makes the area of copyright difficult, especially when it applies to the digital arena of the web where copying is so easy and natural. Here we present a snapshot of the issues for online copyright, with particular emphasis on the relevance to cultural institutions. We concentrate on Europe and the US; as an example we include a special section dedicated to the situation in Italy.**

## INTRODUCTION

Copyright, especially online [3], is a very complicated problem to discuss and solve, although it is an important issue for art galleries and museums in the promotion of their resources [13][17]. There are many different interests to consider, often difficult to bring together, since very frequently they are contradictory with each other. It is a very intricate issue, particularly because it is not clear how we can find a proper answer to the significant questions raised by the mix of protecting traditional intellectual property rights and the development of technology that allows fast efficient distribution of the same content without the intervention of the traditional intermediaries that have organized distribution until now.

It should be understood that there is no unique answer or definitive truth about copyright and in this paper we are trying to raise questions more than to respond to them, in the light of present and future developments of technology. We must also emphasise that the issue cannot be easily synthesized as simply a struggle between legitimate "property" against "piracy". There is much more than that at stake in this battle, since there is the possibility of free culture and creativity in a free market. The copyright topic is of a huge social, economic, legal and technical relevance, because it regards the way digital information is distributed and used in an electronic "area". It goes far beyond the tiny problem of piracy of intellectual property rights for creative content.

## CONSUMERS' CONCERNS

According to the State-of-the-Art Report on Digital Rights Management (DRM) written by the European Consortium INDICARE (The Informed Dialogue about Consumer Acceptability of DRM Solutions in Europe) [7], there are many concerns of consumers with respect to the different tools used for the Digital Rights Management, although most of them are still unaware of the problems under discussion here. The major areas of risks as well as of strong expectations among users when dealing with digital content are:

- Ease of access to and use of content
- Privacy
- Transparency and fair contract terms
- Interoperability
- Security and hardware issues
- Flexibility in business models
- Product diversity and pricing

We will try to address some of these issues in order to clarify the state of the present legislation and the technological solutions for various problems with special regard to multimedia (especially video) content accessible on the Internet. We would like to stress, though, that with relation to the nature of digital content, there is not a great deal of difference conceptually between written, image, audio or video content. An important practical variant is that while with text, image and audio content the bandwidth of the network has mostly already been sufficient for the last few years even with the availability of relatively slow technology, the issue of free download of video content must now be faced with the advent of increasingly fast and widespread broadband connections. So we can gain some hints on the digital video distribution future trends by looking, for instance, at the actual existing distribution habits of music online.

While particular content is essentially fixed in the digital arena, the consumer is not unique. Different people have varying characteristics, needs and expectations. There are in fact many diverse groups of users: librarians, private users, public institution users, educational users, users with disabilities, etc. Each group presents its own concerns with regards to Digital Rights Management. However, all groups of users share a common expectation: they generally consider that all the content uses and accesses that were allowed in the traditional analogue world should also be available also in a digital environment. Consumers are not prepared to accept that the transfer of content in a digital format prevents them from making copies of the content for their personal use, choosing among playback devices, using content whose copyright has expired, using content anonymously, annotating content for personal use, having continuous access and political freedom, making use of the first sale doctrine (allowing personal loan, reselling or giving away), accessing content in libraries, freely copying for didactic purposes, etc. ([7], page 21). They are open to increasing their opportunities as consumers but not to diminishing their rights and having to ask for permission in contexts where the equivalent analogue content was freely allowed previously. Consumers are unlikely to find the controlling attitude of content providers acceptable, contrasting with freer digital production and distribution tools than is increasingly available.

## CONTENT PROVIDERS AND PUBLIC INSTITUTIONS POLICIES

Content providers are experiencing technical impotence in protecting their previous privileges in the digital networked world and the lack of guarantees when using the new distribution channels offered by the network such as Peer to Peer (P2P) networking. Their feelings of insecurity have pushed them to concentrate all their efforts on strong and persuasive lobbying activities towards national and internationals institutions in order to restrict the use of the intellectual property rights as much as possible in their favour. A first successful result of this campaign to protect their rights is the present legislation both in the US (The Digital Millennium Copyright Act [15]) and in Europe (Directive on the Enforcement of Intellectual Property Rights [6]).

If content providers are the most powerful actors on the scene, and their role is very well understood and protected by international and national governmental institutions, other considerations need to be analyzed in order to understand the real potentialities, opportunities and risks of this battle. In this paper, we try to address some of the issues at stake in this discussion.

## PIRACY AGAINST PROPERTY?

We would like to consider the story of the media industry as a story of "piracy" [http://wikisource.org/wiki/Free_Culture_(Piracy)], in which every medium has "stolen" the property rights from previous media, trying to exploit the new opportunities provided by new technologies, such as cinema, radio, and cable television. As Lawrence Lessig [8] noticed, commenting about the parallel between the P2P distribution technology and other media births:

*(1) Like the original Hollywood, P2P sharing escapes an overly controlling industry; and*

*(2) like the original recording industry, it simply exploits a new way to distribute content; but*

*(3) unlike cable TV, no one is selling the content that is shared on P2P services.*

*These differences distinguish P2P sharing from true piracy. They should push us to find a way to protect artists, while enabling this sharing to survive* ([8], page 66).

Technology shapes society in the sense that it offers new ways of looking at the world and new theories about it. Some tools, such as the telescope or the microscope, changed the vision of the world more clearly and directly; in other examples, new technology acts more implicitly and indirectly, but no less deeply, as happened with mass media and "new media". Society always offers resistance to change, because it provokes a redefinition of the borders of reality, which entails a reorganization of powers and influences that cancel some privileges to create a new status quo; however, it is highly unlikely that the old powers can win the battle against technology in the long term.

In the specific case of P2P distribution technology for digital content, it seems apparent that it is important to find a way to protect the legitimate rights of creators, and we have to answer the question about how much money the distribution industry has lost, as a consequence of the widespread use of it. But we also have to ask ourselves other questions, such as the following. How much content would not be available if it were not distributed via P2P networks? How much does society itself gain out of the diffusion of this new tool to share information?

The concept of "original" with respect to copies in the "age of mechanical reproducibility" is a loose one. In fact, there is absolutely no difference between the "original" and the "copies" in the digital format; moreover if you download a copy of a video from a P2P library you are not stealing the copy because the "original" remains there. The duplication does not lose quality or damage the original in any way. The work of art becomes immaterial and, as Walter Benjamin [2] noticed about 70 years ago, speaking about photography and cinema, it loses the "aura" that was involved in the authenticity of the original. According to him, "the technique of reproduction detaches the reproduced object from the domain of tradition." This does not mean that we should cancel intellectual property rights as we knew them before the change to digital technology, but digital reproducibility needs to be tackled from a new perspective in order to imagine a correct, legitimate, adequate and acceptable way to respect people's rights.

According to a study by Ipsos-Insight in September 2002, 60 million people have downloaded music via P2P networks in US (28% of Americans over 12) and according to a survey by NDP cited by the New York Times in May 2003, 43 million people in US used file-sharing technology to exchange material ([8], page 67). We cannot limit ourselves to consider all these people just as insignificant pirates who want to steal copyright-protected content. We must reflect on the phenomenon more deeply and try to interpret the situation in terms of a change of habits and attitude towards digital content.

There is also an interesting recent empirical study (March 2004) that shows that it is not true that the download of music using the Peer2Peer networks creates a loss for the distribution companies (majors) of the music. According to the research, in fact, it seems apparent not only there is no diminishing in the sales but also that P2P file sharing popularity is directly related to the increase in the sales of the corresponding album [10]. Even if this proof were not necessary, because it was intuitive that this was the case, it is evident now that file sharing technology can act as a sort of promotion for music: consumers listen to the file downloading it online and when they are happy with the quality of the music they tend to buy it, even more than before. It is likely also that the file sharing can offer new promotion possibilities for singers that are less famous and/or less sponsored by the major distribution companies, creating a new distribution channel for "art works".

## CREATIVITY AND "BORROWING" FROM PREVIOUS CREATORS

Copyright was created to protect and encourage creativity, but now it seems to be used to block creative efforts of contemporaries to protect the rights of past creators or, more precisely, present administrators of rights of past creators. As underlined by the editorial note in [12], if Shakespeare were working today on Broadway or in London's West End, he would be spending a lot of time dealing with lawyers. Most of his famous successes such as Romeo and Juliet came from other people's ideas that he adapted for his plays. But he is not the only example; the history of artistic and scientific creativity is based on copying and citing works by other artists and scientists. The creative process can be resumed by this citation from one of the most influent philosophers of the last century, Jacques Derrida (1930–2004):

> [...] *je suis très fatigué, pour de raisons qui n'ont aucun intérêt pour vous, mais qui expliquent que je n'aie pas pu préparer un texte qui ressemble à une conférence. Donc, je vais faire ce qu'on a appelé hier de «la rhapsodie». On a beaucoup parlé hier de rhapsode, c'est-à-dire de quelqu'un qui, avant l'existence de la propriété des droits d'auteur, pouvait librement «piquer» un truc là, le coudre, en découdre, sans que personne ne pose de questions* [4].

This citation is not only relevant in the limited context given by Derrida, but can apply to all creative works. The policies, following those of international institutions, of increasing the number of years of copyright protection, while escalating the penalties for copyright infringements, is not a good response to help protect, sustain and develop the creativity of contemporary artists. In the process of protecting past works of art, it seems that present artists are prevented from the possibility of developing and promoting their activity in freedom.

In the next section, we consider the legislative situation in Italy with respect to copyright as a specific example, especially since there have been some signification changes in that country recently.

## COPYRIGHT PROTECTION IN ITALY: THE NEW LEGISLATION

Between March and May 2004 the Italian Government decided suddenly that the copyright law, which had been in place since 1941 – established during the fascist era and in force for 63 years, almost without modification – needed to be revised. The Italian Government used a special procedure, the "Decreto Legge" (Decree) that is allowed by the Constitution only when there is a clear necessity and urgency of the reorganization of the rules. The copyright issue did not seem so vital to everyone, but the government decided otherwise. There was apparently no pressure to define a new copyright protection system, but the tool used allowed the Government to define all the details of the law and to control the discussions in both the Parliament and the Senate. Once the new law was proposed using that special procedure, it was necessary to approve it, converting it into a real law within 60 days. The "Decreto Legge" (DL 22/3/2004 N. 72), approved in March 2004 from the Government, was transformed into a Law of the State (L. 21/5/2004 N. 128) in May 2004 [11], with the promise of making further changes in the near future, because, due to the rapidity of the approval process, it was clear to everybody, even to the MPs who approved it, that there were many mistakes, problems and even risks of permitting unconstitutional rules.

It is still too early to draw conclusions about the consequences of the reorganization. However we can describe the solutions chosen and the potential inconveniences created by the new situation.

One of the major changes of the new legislation is the change of the degree of the copyright infringement offence. Before 2004 in Italy the offence was committed only if the copy of the protected material was realized for a direct lucrative objective. Now it is considered an offence the simple copy for a generic profit aim. So according to the law, if you make an electronic copy of a CD that you bought, paying a fair price, and you keep it in your hard disk while belonging to a Peer2Peer network, the bare fact that you allow other people to download the electronic copy of your file, giving you the opportunity to make a profit of it in exchange, means that you have committed an offence. This is true also if you look only for material that is already free from copyright protection, because you can achieve a profit anyway.

The new formulation of the offence is connected with another relevant change: the criminalization of the use of certain technologies such as some electronic distribution methods: Peer2Peer. The idea is that you commit an offence if you simply "use any telecommunication system to distribute an entire work or a part of it that is protected by the copyright law" ([11], Art.1.3). The generic profit aim includes all sort of non-lucrative enterprises such as libraries, research institutions, universities, museums, etc.

that create a data bank of electronic material, or even a website in which it is possible to access some of the content (protected by the copyright law) that is in their possession. The aim of such institutions should be to protect the right of the users to access as much information as possible and increase the awareness of the available content among their users. However this behaviour represents a so-called "profit" for the cultural institutions in terms of recognition of their authorities and increase of their "brand awareness" among users that could be exploited if they decided to organize an event or an exhibition, etc.

"Fair use" by such institutions is not protected or even mentioned in the legislation, while it is clear that everyone can make copies of legally owned material for personal use. However, libraries sometimes need to make copies of material for social use, but the law does not guarantee permission for their activities, while pursuing the social role that is clearly stated by their legal identities and their natural objectives.

The Italian law must be in line with European legislation and it should be noticed that there is a discrepancy here between the lack of consideration of the rights of the users and the strong protection of the rights of the producers and of the distributors of "art works". Instead, according to the European Directive 2001/29/EC:

> *A fair balance of rights and interests between* […] *the different categories of rightholders and users of protected subject matter must be safeguarded* [5].

In case of inconsistency between the national and the European legislation, an Italian judge should be obliged to apply the European principle. In the European Directive, however, it is not very clear by what means by which the users' rights could be "safeguarded". In a situation of legal uncertainty the weakest people tend to be penalized, because they do not have the financial backing to defend themselves before the law using all the available tools. In this case we can affirm that the cultural institutions are likely to be more careful than before in launching digitalization initiatives that involve dealing with copyright protection issues and their prudent behaviour will result in a limitation of the end users' rights.

In order to monitor suspect copyright infringements, Internet Service Providers (ISP) are charged with especially onerous duties: if the judiciary authority asks for their cooperation in the investigation of potential offences, they are obliged to provide all the useful information about their clients who could be guilty of copyright infringements. This obligation is severe and they risk a very expensive fine in case of lack of action (up to €250,000). The technological effort in order to guarantee observation of the behaviours of all their clients is particularly arduous and expensive, especially for small ISPs, such as universities and other institutions that offer Internet service without asking for payments from their clients. Moreover, the delicate issue of the privacy of clients has to be taken into account too. ISPs are thus in a dilemma: on one hand they are supposed to respond efficiently to legal requests; on the other hand they must guarantee the respect of the free expression of thought and the protection of the other major liberties of private citizens, guaranteed by the Constitution.

There is also another strategic change in the copyright protection principle: the transformation of infringement into a criminal offence that allows the judge to sentence a convicted person with up to four years of imprisonment. According to the previous law, this was only an administrative misdemeanour that could only be punished with a fine. This is not a minor question, because it is related to the category of offence and consequently to the ontology of the juridical principle infringed in committing the offence. In order for the behaviour to be considered criminal, a significant social danger should be involved; however the social menace at stake in copyright infringement activities in unclear.

The last issue that is raised by the new law relates to the institution of a new obligation: the adequate demonstration of a clear statement that there is no infringement of the copyright law, whenever much electronic material is published. This is a very complicated request that is not well defined and is another area of ambiguity of the Decree. It risks generating misunderstandings and problems for users as well as ISPs that are responsible for checking that this is the case for all the content made available using their network facilities (including websites, email, weblogs, etc.).

The Italian situation is at the moment effectively in a state of limbo, because everybody knows that the law cannot be applied completely due to inconsistencies and confusions. It is not easy to advance a new law on the same subject so quickly, though it was explicitly promised by the Government during the approval of the "emergency measure". The only hope is that the sensitivity of the subject and of the cultural issues will succeed in producing a proposal that allows a radical change in the balance of the rules for more reasonable copyright protection that is more respectful of users' rights.

## PROPOSED SOLUTIONS AND FUTURE REFLECTIONS

The problems discussed above can be solved in different ways according to the issues at stake. We can work for a technological solution, a business solution or a legal solution, for example.

The technological solution would imply the control of digital content via security and encryption systems, or tagging and watermarking content. However all these solutions are never completely guaranteed to ensure security, as can be seen from the history of "cracking" all sorts of cryptographic protection systems that have been designed to prevent copying and conversion from one format to another. Sometimes, in terms of promotion of content, it is more useful to allow its distribution in order for it to become more well-known and desirable for subsequent acquisition in protected formats.

The business solution presupposes the invention of new business models that make it more convenient for the consumer to buy the rights to the content than to find the way of cracking it via the web. These new business models could easily use the same distribution channels offered by the network such as Internet providers or the iTunes selling model [http://www.itunes.com]. Another method that is likely to be used to distribute digital content is the use of sponsorship and advertising models (the same business model as commercial television), which are particularly interesting for entertainment video content.

There is also the distribution of content for marketing and content promotion; this model is already used for movie trailers, for example. This model could also be used to promote complementary product and services, such as mobile or broadband services. Video content can be considered as a free companion to the officially sold service. A special case of this model is the syndication of content to portals or other service providers that buy the content (text, video or audio) to make it available for no cost or at a low rate to customers, using it to promote their primary service ([7], pages 104–107).

There are various business models that can be foreseen for selling digital content using particular distribution channels or as a stand-alone business model. It is important to bear in mind that only creative business models will be able to cope with the new distribution networks and compete with free file-sharing distribution. Only if customers perceive the added value of buying a product that can be easily accessible at zero cost will they agree to pay for it.

The third level of solution is the legal one. It must be clear that it is unbalanced only to take into account the rights of creators (or of the mediators, such as the traditional distribution companies that normally control those rights, especially for video content). There are also the rights of the consumers to access the content easily. We have library rights to enable them to distribute content among users, third world countries consumers that do not have the economic power to buy the digital content at the same price as western world consumers, consumers with disabilities that have special rights to access content, educational institutions rights for student use of content, etc.

A way to cope with all these different expectations was suggested in 2001 by Lawrence Lessig, who became the leading force behind Creative Commons (CC) [http://creativecommons.org], a non-profit group that offered a legal way to oppose to the lack of balance of Intellectual Property protection with respect to the "all rights reserved" license. Creative Commons proposes all sorts of free licenses by which authors can decide to give away all or some of their copyright privileges. One license permits others to use a work as long as the author is correctly mentioned, another gives the right to sample as long as the entire work is not used, another allows the giving away of the content for some third world countries citizens, and so on.

There are currently more than 5 million CC licenses in use around the world. The BBC is planning to license archival material to the British public without fee, as long as it is only used for private purposes [1]. MIT used the license for giving open public access to "OpenCourseWare" online course materials [http://ocw.mit.edu] and there is also a Science Commons [http://science.creativecommons.org] that is trying to explore the possibility of offering a free license for some of the content of scientific patents.

There are various interesting and promising initiatives also in the field of video archiving and content distribution. One of the major actors in the area is the *Internet Archive* [http://www.archive.org]. They host various projects under a Moving Image Archive [http://www.archive.org/details/movies], including the Prelinger Archive, a freely accessible ephemeral video archive of over 48,000 films founded by Rick Prelinger in 1983, as well as open source movies, etc. We hope that in the future these initiatives will not remain isolated examples of promotion of open access to digital content, but will be followed by cultural institutions that have, as part of their remit, the spread of information in order to facilitate the creation of knowledge.

The copyright debate related to digital content is still only at the beginning at the moment. There are various powers facing each other, each seeking a convenient solution for their own interests. The legislative side of the story currently revolves around lobbyists for major production and distribution companies, but this is still at an early stage. At the US Senate in 2002, a new Bill (the Digital Media Consumers' Rights Act [14][16]) has been proposed; this would offer a more balanced solution to the digital content copyright that defends the expectations of consumers better with respect to the Intellectual Property rights protection. The future is still confused and unknown, but looking back at the history of the media industry, technological innovations have always brought a change in the organization of distribution that in the end favours innovators rather than conservators. The process of change will take time, because it is a struggle between old and new media, but it is very unlikely that Internet users will have to renounce to the new opportunities of file-sharing distribution technologies to obtain content.

## USEFUL LINKS

- Creative Commons
  http://creativecommons.org
- Free Software Foundation [18], US & Europe
  http://www.fsf.org & http://www.fsfeurope.org
- Intellectual Property, UK Government
  http://www.intellectual-property.gov.uk
- JISC Legal Information Service – IPR Useful Links
  http://www.jisclegal.ac.uk/ipr/IntellectualPropertyLinks.htm
- Motion Picture Association of America – Anti-piracy
  http://www.mpaa.org/anti-piracy
- Respect Copyrights, Illegal Trafficking in Movies
  http://www.respectcopyrights.org
- Stanford Copyright & Fair Use Center
  http://fairuse.stanford.edu
- UK Patent Office – Copyright
  http://www.patent.gov.uk/copy
- What is Copyright Protection?
  http://whatiscopyright.org
- WIPO: World Intellectual Property Organization
  http://www.wipo.int

## ACKNOWLEDGEMENTS

Dr. Teresa Numerico was a Visiting Leverhulme Research Fellow at London South Bank University from June 2004 to April 2005, when most of this research was undertaken. She has since returned to a permanent academic post at the University of Salerno, Italy. This paper has been expanded from a presentation originally given as part of the *Film on the Web* conference at the National Museum of Photography, Film and Television, Bradford, UK [9].